# Analysis Acoustic Features for Acoustic Scene Classification and Score fusion of multi-classification systems applied to DCASE 2016 challenge

*Sangwook Park* [1], *Seongkyu Mun* [2], *Younglo Lee* [1], *David K. Han* [3], *and Hanseok Ko* [1]

[1] School of Electrical Engineering, Korea University, Seoul, South Korea
[2] Department of Visual Information Processing, Korea University, Seoul, South Korea
[3] Office of Naval Research, Arlington, VA, USA

{swpark, skmoon, yllee}@ispl.korea.ac.kr, ctmkhan@gmail.com, hsko@korea.ac.kr

## Abstract

This paper describes an acoustic scene classification method which achieved the 4[th] ranking result in the IEEE AASP challenge of Detection and Classification of Acoustic Scenes and Events 2016. In order to accomplish the ensuing task, several methods are explored in three aspects: feature extraction, feature transformation, and score fusion for final decision. In the part of feature extraction, several features are investigated for effective acoustic scene classification. For resolving the issue that the same sound can be heard in different places, a feature transformation is applied for better separation for classification. From these, several systems based on different feature sets are devised for classification. The final result is determined by fusing the individual systems. The method is demonstrated and validated by the experiment conducted using the Challenge database.

## 1. Introduction

Audio signal including speech, general sound (non-linguistic sound), and background sound may be quite informative in characterizing context such as presence of humans, objects, their activities, or the environment. Among these, location information is not only vital in multimedia analysis but also used in many tasks of obtaining clues that are helpful in scene understanding. Location information is widely used in many applications [1-2] including speech/acoustic event recognition as a prior for enhancing the performance [3-4]. Thus, Acoustic Scene Classification (ASC) that recognizes the location where the observation is obtained has drawn considerable attention due to these usages.

Typically, ASC system consists of feature extraction and classification. In feature extraction, observation data having discriminative characteristics depending on locations are expressed as a feature vector. Mel Frequency Cepstral Coefficients (MFCCs) and Perceptual Linear Prediction (PLP) have been applied in early ASC system. Low-level spectral features such as zero-crossing rate, spectral statistics, and timbre were employed for ASC by combining them with MFCCs [5]. Chu [6] utilized a joint feature between MFCCs and time-frequency features obtained from the matching pursuit decomposition of acoustic scenes to improve the overall performance. In other approaches, a feature vector is extracted by considering variations among the frames. A Bag-Of-Frames (BOF) method, which considers an acoustic scene as a set of bags of various sounds, was applied to ASC [7-9]. This approach used statistical distribution (e.g. histogram) as features, which represents the occurrence count of cepstral features, quantized by a codebook like dictionary. In classification, location where the observation is obtained can be revealed by a discriminant function which is obtained based on Gaussian Mixture Model (GMM) or Support Vector Machine (SVM) [1, 7].

Although many approaches were applied for ASC applications, these still suffer from problems in realistic environments. Even in a same place, a variety of acoustic sounds occur depending on presence of people, objects, and their behaviors. For an example, when a microphone is set in a café, different types of sound can be captured depending on happenings such as cleaning, coffee grinding, or people talking. As such, feature vectors obtained in the café can be widely scattered in a feature space, although these vectors come from the same place. On the other hand, conversations can well be heard where ever there are people. As such, the feature vectors obtained from different locations may well be very close to each other. As illustrated by the example here, ASC for location classification is still very challenging in realistic environments.

The main contributions of this paper are to establish an improved framework and a method for ASC as follow: 1) analysis of signal components composing an acoustic scene signal, and investigation of several features in terms of their components, 2) investigation of a novel feature and feature transformation for resolving the issue of location classification from the same sound source in different locations, and 3) development of a fusion method based on each system output for improving the overall ASC performance.

The remainder of this paper is organized as follows. Section 2 explains the approach with its motivation. Section 3 introduces ASC approaches. After a discussion on the experimental results, conclusions are drawn in the final section.

## 2. Audio Signal Components

All sounds including a background noise and reverberation can be a signal for ASC, because all of them are helpful for obtaining the location information. An input sound can be categorized into three components: stationarity, quasi-stationary, and non-stationary according to time duration.

As shown in figure 1, an engine sound of a bus can be obviously considered as a stationary component. In case of break noise of a bus, however, these sounds are considered as a quasi-stationary component because it can be considered as a stationary for a moment. And many sounds that suddenly happens are frequently captured in a short period of time. Thus, they are considered as non-stationary.

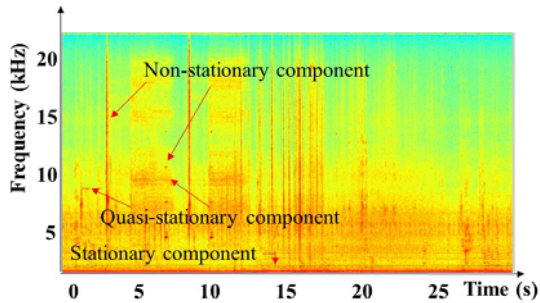

Figure 1: *Sample spectrogram obtained in a "bus" that is released by the DCASE 2016.*

A stationary component is the primary factor for a target locational information. The others are secondary factors because most of these acoustic events depend on people/entities in the location, and their behavior.

## 3. Approaches for Acoustic Scene Classification

The system proposed here is depicted in figure 2 with three main parts; Feature Extraction (FE), Classification (CL), and Score Fusion (SF). The methods applied in the system are described in the following.

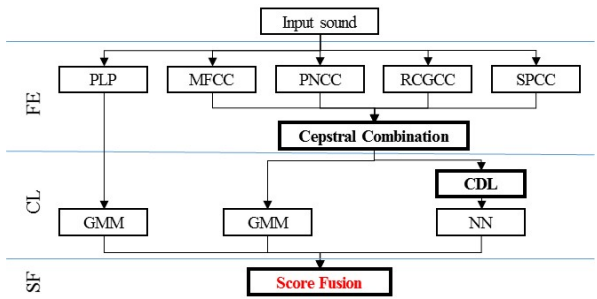

Figure 2: System architecture applied in DCASE 2016

### 3.1. Feature Extraction (FE)

#### 3.1.1. PLP and MFCC

MFCC and PLP have shown good performance in recognition. In ASC, both features are obtained under the condition that all of the three stationarity-based components (stationary, quasi-stationary, non-stationary) are mixed. In the case of PLP, an equal loudness pre-emphasis is applied to approximate the result of frequency integration to the inequivalent sensitivity of human auditory sense at different frequencies [10]. Thus, discriminative characters in high frequency band are included in the process although the energy contribution in high frequency is relatively low due to their rapid attenuation.

#### 3.1.2. PNCC

The procedure for the Power Normalized Cepstral Coefficients (PNCCs) called Medium-time power bias subtraction is added in the signal enhancement method [11]. Since the stationary component is considered as a power-bias in the procedure, the feature is extracted under the condition that the stationary component is suppressed.

#### 3.1.3. RCGCC

The Robust Compressive Gamma-chirp filterbank Cepstral Coefficients (RCGCCs) was added as signal enhancement [12]. In our implementation of RCGCC the features are extracted by suppressing a stationary component as in the case with the PNCC. Additionally the non-stationary components are enhanced by applying a smoothed weight to the observation subband power.

#### 3.1.4. SPCC

Since many sounds are non-stationary and highly dynamic, feature vectors extracted in a location may be widely scattered in the feature space. For resolving the issue, Subspace Projection Cepstral Coefficients (SPCC) is applied [13]. The rank of each subspace is determined by preserving 90% of data energy that is a summation of all eigenvalues.

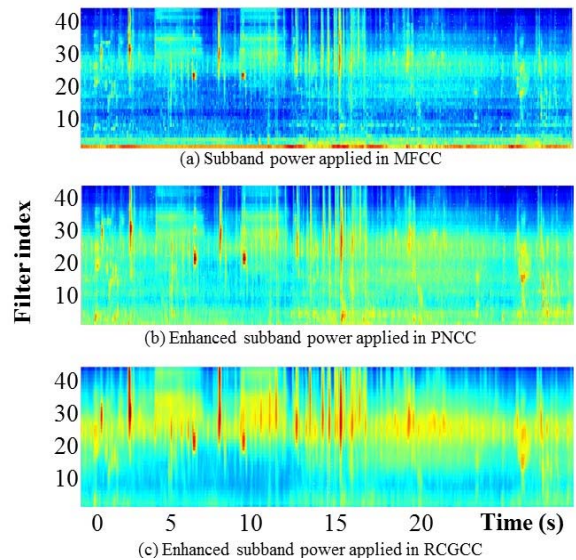

Figure 3: Subband powers applied to each feature extraction method

#### 3.1.5. Cepstral Combination

For summary of previously mentioned features, figure 3 shows several subband powers that are intermediately obtained in each feature extraction using an observation depicted in figure 1. A subband power applied to MFCC is shown in figure 3(a) that is similar to the spectrogram representing the mixed components. Enhanced subband powers for PNCC and RCGCC is shown in figure 3(b) and figure 3(c), respectively. In these cases, a stationary component is diminished. In particular, the high frequency bands corresponding to non-stationary component are more noticeable in the enhanced subband power for RCGCC. Whether quasi-stationary component will be reduced or not is determined by its time duration.

Since sounds are continuously changing according to situations such as human behaviors and other sound sources, feature vectors extracted in every frame are widely spread in vector space, even though the features are obtained from a common place. In order to complement this issue, a novel feature is also applied. According to advantages of each method, a Cepstral Combination (CepsCom) that is made up of concatenating four features; MFCC, PNCC, RCGCC, and SPCC are used for ASC.

Table 1. *Accuracies of experiment using Development dataset; the number of mixture is set to 4 and 64 in PLP and others except CepsCom-CDL (denoted by "Ceps-CDL") and Fusion, respectively.*

| [%] | Avg. | beach | bus | cafe | car | city | forest | groc. | home | lib. | metro | office | park | resid. | train | tram |
|---|---|---|---|---|---|---|---|---|---|---|---|---|---|---|---|---|
| MFCC | 66.83 | 66.67 | 68.80 | 64.96 | 64.96 | **85.90** | 59.83 | 63.25 | 72.22 | 69.66 | 63.68 | **71.37** | 56.84 | 66.24 | 47.86 | 80.34 |
| PNCC | 63.59 | **69.23** | 50.85 | **77.35** | 68.80 | 79.49 | 52.14 | 66.67 | 54.27 | 71.37 | 62.82 | 60.26 | **57.26** | 64.10 | 46.58 | 72.65 |
| RCGCC | 63.65 | 62.82 | 55.98 | 64.10 | 64.53 | 84.62 | 55.56 | 75.21 | 50.43 | 70.94 | 63.25 | 62.82 | 53.85 | 61.11 | 48.72 | 80.77 |
| SPCC | **71.51** | 58.97 | **82.05** | 60.68 | **82.05** | 83.76 | **61.97** | **76.92** | **82.48** | **72.22** | **81.62** | 70.51 | 50.85 | **76.07** | **51.28** | **81.20** |
| CepsCom | 73.99 | 70.09 | 74.79 | **72.65** | 85.04 | 88.89 | 61.97 | **86.32** | 76.92 | 71.37 | 76.07 | 74.79 | 55.13 | 75.64 | 50.00 | **90.17** |
| Ceps-CDL | 74.62 | 70.51 | 78.21 | 49.15 | 89.74 | **91.88** | 79.06 | 85.04 | 63.68 | **83.33** | 77.35 | **82.48** | **61.97** | 73.50 | 50.00 | 83.33 |
| PLP | 68.43 | 63.25 | 73.50 | 51.71 | 73.50 | 76.50 | **86.75** | 72.65 | **79.91** | 57.26 | 75.64 | 54.70 | 59.83 | 68.38 | 45.73 | 87.18 |
| Fusion | **76.36** | 68.80 | 75.64 | 61.54 | **91.88** | 85.04 | 85.90 | 79.49 | 79.91 | 79.91 | **90.60** | 76.07 | 49.57 | **78.21** | **57.69** | 85.04 |

### 3.2. Feature Transformation and Classification

To resolve the issue that the same sound source can be heard in many different places, feature transformation is considered. [t6] Although the same sound in different locational settings can be closely mapped in a feature space, their feature distributions may be different enough to distinguish if they are observed in different places. R. Wang's method of Covariance Discriminative Learning (CDL) as a transformation is implemented here [14].

In classification, GMM is considered for frame based features including PLP and the other cepstral features. Nearest Neighbor (NN) is used for the result of CDL.

### 3.3. Score Fusion

Accuracies of each class may be different depending on the approach. Thus, overall averaging accuracy can be improved by considering all outputs from several systems. AND, OR, and Majority Vote are some of the commonly used fusion rules. However, these approaches have equal confidences for all outputs from each individual system, though one of the systems may show more reliable output for a certain class. The approach proposed here is by way of weighted fusion as the final result computed in (1) as

$$final\_result = \arg\max_c \left( \sum_{n=1}^{N} w_c^n s_c^n \right) \quad (1)$$

where $c$ and $n$ are indexes about class and individual system. $s_c^n$ and $w_c^n$ is a score normalized into the interval [0, 1] and its reliability (i.e. weight) in each individual system. In GMM and NN, likelihood and distance margin are defined as the score, respectively. $N$ is the number of individual systems applied for score fusion. The weight $w_c^n$ is defined as in

$$w_c^n = P(G=c | O=c) = \frac{P(G=c, O=c)}{\sum_g P(G=g, O=c)} \quad (2)$$

where $G$ and $O$ are random variables meaning a ground truth and an output of an individual system, respectively. The more outputs are confused with the class $c$, the lower the weight is applied to its score, even if the class $c$ is perfectly recognized in an individual system. The weight can be determined from a confusion matrix obtained from the training data used for each individual system.

## 4. Experiments

### 4.1. Database and Experiment setting

For performance assessment, the proposed system was performed with the database released for DCASE challenge 2016 [15]. The database consists of development and evaluation dataset. Both include 15 acoustic scenes: namely *bus*, *café/restaurant*, *car*, *city center*, *forest path*, *grocery store*, *home*, *lakeside beach*, *library*, *metro station*, *office*, *residential area*, *train*, *tram*, and *urban park*.

Firstly, performance assessment was conducted by using only the Development dataset. In this case, training and test sets were assigned with a ratio of 1:3, because the amount of training case is limited in comparison to the test case in a real situation. In the second experiment, Development and Evaluation datasets were used for training and test, respectively. Note that the weight used for score fusion is determined in the training.

For feature extraction, frames were defined as 2048 samples with an overlap with the next frame for 1024 samples. Discrete Fourier transform was conducted with a 2048 points Hamming window. In case of PLP, 39 coefficients including delta, acceleration, and energy were extracted by using HTK [16]. In the others; MFCC, PNCC, RCGCC, and SPCC, 60 coefficients including delta and acceleration were extracted. In this case, CepsCom is a 240-dimensional vector.

### 4.2. Experiment results and Discussions

#### 4.2.1. Development of ASC system

The accuracies according to classes are summarized in Table 1. Among the cepstral features, SPCC shows the best accuracy in most of the classes. In case of 5 classes, MFCC or PNCC is better than SPCC, and RCGCC shows the second best in *city*, *grocery*, *train*, and *tram*. An average accuracy of CepsCom shows better than the other cepstral features. Especially, class accuracy outperforms cepstral features that are elements of the CepsCom in *beach*, *car*, *city*, *grocery*, *office*, and *tram*, because the distance between confusable features is sufficiently extended to distinguish the features by expanding a feature space. Note that the PLP is excluded from forming CepsCom because accuracy variation depending on the number of mixture is somewhat different to other cepstral features [17].

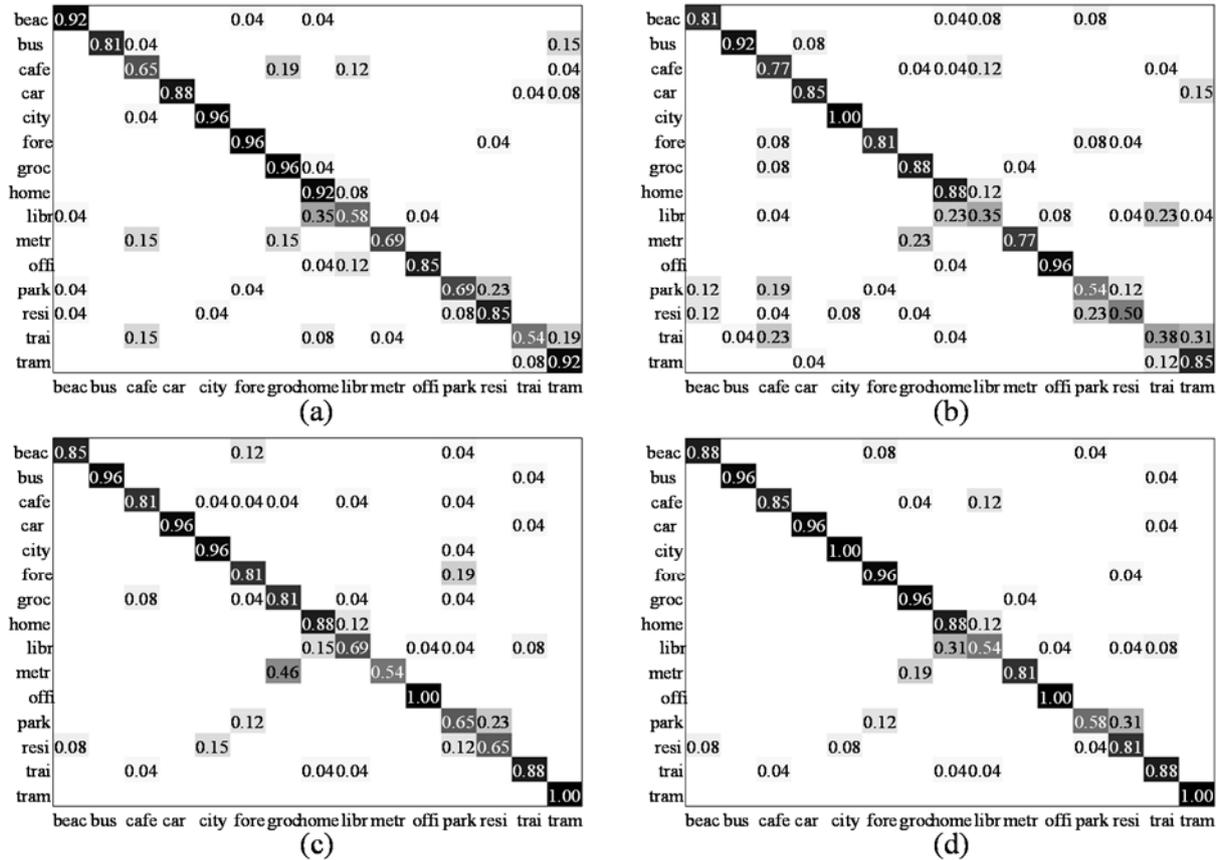

Figure 4: Confusion matrices; (a) PLP-GMM (b) CepsCom-GMM (c) CepsCom-CDL (d) Fusion

When the CDL is applied, the CepsCom shows the best result among all considered features. Although the average accuracy of CepsCom-CDL is similar to CepsCom-GMM, each class accuracy is different to each other as shown in Table 1. Also, PLP-GMM system shows a better accuracy in forest and home. Based on this fact, CepsCom-GMM, CepsCom-CDL, and PLP-GMM are used to make the final decision. The result of the fusion shows the best average accuracy among individual systems. In particular, accuracies of car, metro, residential, and train are better than individual systems applied to the fusion. Since a low weight for unreliable output from an individual system is applied to the fusion, the best candidate after the fusion may well be the second best candidate in an individual system.

*4.2.2. Evaluation of ASC system*

From the previous experiment, three systems were evaluated with Evaluation dataset. In results, average accuracies of all individual systems are much better than the previous one, because the number of training data is increased as much as four times. Confusion matrices of each individual system are summarized in figure 4. As shown in the figure, the PLP-GMM especially shows more than 95% accuracy in *city*, *forest,* and *grocery,* while three groups; *home-library*, *park-residential*, and *train-tram* are confused to each other. Although the CepsCom-GMM also suffers from confusions in these groups, the system shows better accuracies than PLP-GMM in *bus*, *café*, *city*, *metro*, and *office*. The CepsCom-CDL also shows better accuracies than PLP-GMM and CepsCom-GMM in several classes, and confusion problem *train* and *tram* is resolved.

In the result of fusion method, the average accuracy is shown as 87.18%, which is an improvement of 5.90%, 12.05%, 6.10% compared to PLP-GMM, CepsCom-GMM, and CepsCom-CDL, respectively. As shown in figure 4(d), two groups; *home-library* and *park-residential* are still confused, but *train-tram* confusion is resolved by the CepsCom-CDL. In several classes, accuracy is improved after applying the fusion method. In particular, accuracies of several classes are better or equal compared to the best accuracy among individual systems in *café*, *car*, *metro*, and *train*.

## 5. Conclusions

This paper described the ASC method that ranked the 4[th] in the first task of the IEEE AASP Challenge: DCASE 2016. Features such as PLP, MFCC, PNCC, RCGCC, and SPCC are investigated, and these features are applied for scene classification by means of GMM for performance evaluation. Also, CDL is applied for feature transformation. From the fact that each class has different accuracy according to classification approaches, score fusion method is developed for making the final decision. In the experiment, the challenge database is used for performance assessment, and the results are summarized and discussed to demonstrate these approaches applied to DCASE challenge.

# 6. References

[1] W. Choi, S. Kim, M. Keum, D. K. Han, and H. Ko, "Acoustic and visual signal based context awareness system for mobile application," *IEEE Trans. Consum. Electron.*, vol. 57, no. 2, pp. 738-746, 2011.

[2] K. Yamano and K. Itou, "Browsing audio life-log data using acoustic and location information," in *IEEE Int. Conf. Mobile Ubiquitous Computing, Systems, Services and Technologies*, pp. 96-101, 2009.

[3] T. Nishiura, S. Nakamura, K. Miki, and K. Shikano, "Environmental sound source identification based on hidden markov model for robust speech recognition," *EUROSPEECH* 2003, pp. 2157-2160, 2003.

[4] T. Heittola, A. Mesaros, A. Eronen, and T. Virtanen, "Context-dependent sound event detection," *EURASIP Journal Audio, Speech, and Music Processing*, pp. 1-13, 2013.

[5] A. J. Eronen, V. T. Peltonen, J. T. Tuomi, A. P. Klapuri, S. Fagerlund, T. Sorsa, G. Lorho, and J. Huopaniemi, "Audio-based context recognition," *IEEE Trans. Audio, Speech, and Language Processing*, vol. 14, no. 1, pp. 321-329, 2006.

[6] S. Chu, S. Narayanan, and C. J. Kuo, "Environmental sound recognition with time-frequency audio features," *IEEE Trans. Audio, Speech, and Language Processing*, vol. 17, no. 6, pp. 1142-1158, 2009.

[7] V. Carletti, P. Foggia, G. Percannella, A. Saggese, N. Strisciuglio, and M. Vento, "Audio surveillance using a bag of aural words classifier," in *IEEE Int. Conf. on Advanced Video and Signal Based Surveillance*, pp. 81-86, 2013.

[8] J.-J. Aucouturier, B. Defreville, and F. Pachet, "The bag-of-frames approach to audio pattern recognition: A sufficient model for urban soundscapes but not for polyphonic music," *The Journal of the Acoustical Society of America*, vol. 122, no. 2, pp. 881-891, 2007.

[9] S. Pancoast and M. Akbacak, "Bag-of-Audio-Words approach for multimedia event classification," in *INTERSPEECH 2012*, pp. 2105-2108, 2012.

[10] H. Hynek, "Perceptual linear predictive (PLP) analysis of speech," *The Journal of the Acoustical Society of America,* vol. 87, no. 4, pp. 1738-1752, 1990.

[11] C. Kim and R. M. Stern, "Feature extraction for robust speech recognition using a power-law nonlinearity and power-bias subtraction," in *INTERSPEECH 2009*, pp. 28-31, 2009.

[12] M. J. Alam, P. Kenny, and D. O'Shaughnessy. "Robust feature extraction based on an asymmetric level-dependent auditory filterbank and a subband spectrum enhancement technique," *Digital Signal Processing*, vol 29, pp. 147-157, 2014.

[13] S. Park, Y. Lee, D. K. Han, and H. Ko, "Subspace projection cepstral coefficients for noise robust acoustic event recognition," in *IEEE International Conference on Acoustics, Speech and Signal Processing*, pp. 761-765, 2017.

[14] R. Wang, H. Guo, L. S. Davis, and Q. Dai, "Covariance Discriminative Learning: A Natural and Efficient Approach to Image Set Classification," in *IEEE Computer Vision and Pattern Recognition* 2012, pp. 2496-2503, 2012.

[15] A. Mesaros, T. Heittola, and T. Virtanen, "TUT database for acoustic scene classification and sound event detection," in *European Signal Processing Conference* 2016, pp. 1128-1132, 2016.

[16] *The HTK book Version 3.4*, Cambridge University Engineering Department, 2009.

[17] S. Park, S. Mun,, Y. Lee, and H. Ko, "Score fusion of classification systems for acoustic scene classification," *Technical report in Detection and Classification of Acoustic Scenes and Events 2016.*